\documentclass[pra,twocolumn,tightenlines,showpacs,nofootinbib]{revtex4}
\usepackage{bm,dcolumn,amsmath,graphicx,amsfonts,amssymb}

\begin{document}
\title{Calculation of parity non-conservation in xenon and mercury}

\author{V. A. Dzuba and V. V. Flambaum}
\affiliation{School of Physics, University of New South Wales, Sydney 2052,
Australia}

\date{ \today }

\begin{abstract}
We use configuration interaction technique to calculate parity
non-conservation (PNC) in metastable Xe and Hg [proposal of the experiment in 
L. Bougas {\it et al}, Phys. Rev. Lett. {\bf 108}, 210801
(2012)]. Both, nuclear 
spin-independent and nuclear spin-dependent (dominated by the nuclear
anapole moment) parts of the amplitude are considered. The amplitudes
are strongly enhanced by proximity of the states of opposite parity. 
\end{abstract}
\maketitle

\section{Introduction}

The study of the parity non-conservation in atoms is a low-energy,
relatively inexpensive alternative to high-energy search for new
physics beyond the standard model (see, e.g.~\cite{Ginges,DK04,DF12}).
For example, parity non-conservation in cesium is currently the best
low-energy test of the electroweak theory~\cite{Ginges,DF12}. 
It is due to high accuracy of the measurements~\cite{Wood} and its
interpretation~\cite{CsPNC} (see also ~\cite{CsPNCold}) . Since the cesium result is unlikely to be
significantly improved, the focus of the PNC study in atoms has
shifted to two important areas: (i) the PNC measurements for a chain
of isotopes and (ii) the measurements of nuclear anapole moment (see,
e.g.~\cite{DF12}). Most of current or planed PNC experiments in atoms
consider both possibilities. 

The experiments are in progress at Berkeley for Dy and
Yb atoms~\cite{DyPNC,YbPNC}, at TRIUMF for Rb and Fr
atoms~\cite{RbAM,FrPNC}, 
 and at Groningen (KVI)
for Ra$^+$ ions\cite{KVI}. There is an an interesting recent suggestion to
measure  PNC in metastable Xe and Hg~\cite{XeHg}. In the present paper
we support this proposal by the atomic calculations. 

The advantages of using Xe and Hg for the measurements of PNC in a
chain of isotopes and the measurements of anapole moments include: (i)
large number of stable isotopes of both atoms (maximal difference in the number of neutrons is $\Delta N=12$ for Xe
and $\Delta N=8$ for Hg); (ii) presence of two stable isotopes for
each of the atoms with non-zero nuclear spin ($^{129}$Xe, $I=1/2$;
$^{131}$Xe, $I=3/2$; $^{199}$Hg, $I=1/2$; $^{201}$Hg, $I=3/2$), these
isotopes are suitable for the anapole moment measurements; (iii) The
PNC amplitudes in the Xe and Hg atoms are enhanced due to the high nuclear
charge and strong mixing with close states of opposite parity. 

An extra advantage comes from the fact that xenon and mercury nuclei
with non-zero spin have valence  neutron, therefore the nuclear anapole measurements will provide the strength constant for  the neutron-nucleus PNC potential.  The anapole moment so far was measured only for
$^{133}$Cs~\cite{Wood} which has  valence  proton. The data for the xenon and mercury would be
complimentary to those obtained for cesium.

In the present work we use the configuration interaction (CI) technique to
calculate the nuclear spin-independent PNC amplitudes 
caused by the weak nuclear charge and the nuclear spin-dependent PNC
amplitudes dominated by the nuclear anapole moments. The result is
presented in a convenient form as a sum of two contributions for
different hyperfine structure components. This would allow to extract
both the value and the sign of the anapole moments by comparing the
measured amplitudes with the calculated ones.

\section{General formalism}

The Hamiltonian describing parity-nonconserving electron-nuclear
interaction can be written as a sum of the nuclear-spin-independent (SI) and
the nuclear-spin-dependent (SD) parts (we use atomic units: $\hbar = |e| = m_e = 1$):
\begin{eqnarray}
     H_{\rm PNC} &=& H_{\rm SI} + H_{\rm SD} \nonumber \\
      &=& \frac{G_F}{\sqrt{2}}                             
     \Bigl(-\frac{Q_W}{2} \gamma_5 + \frac{\varkappa}{I}
     {\bm \alpha} {\bm I} \Bigr) \rho({\bm r}),
\label{e1}
\end{eqnarray}
where  $G_F \approx 2.2225 \times 10^{-14}$ a.u. is the Fermi constant of
the weak interaction, $Q_W$ is the nuclear weak charge,
$\bm\alpha=\left(
\begin{array}
[c]{cc}%
0 & \bm\sigma\\
\bm\sigma & 0
\end{array}
\right)$ and $\gamma_5$ are the Dirac matrices, $\bm I$ is the
nuclear spin, and $\rho({\bf r})$ is the nuclear density normalized to 1.
The strength of the spin-dependent PNC interaction is proportional to
the dimensionless constant $\varkappa$ which is to be found from the
measurements. There are three major contributions to
$\varkappa$ arising from (i) electromagnetic interaction of atomic
electrons with the nuclear {\em anapole moment} ~\cite{anapole}, (ii) electron-nucleus
spin-dependent weak interaction ~\cite{NSIweak}, and (iii) combined effect of
spin-independent weak interaction and magnetic hyperfine interaction ~\cite{NSDhyperfine}
(see also  ~\cite{Ginges}). In this work we do not distinguish
between different contributions to $\varkappa$ and present the results
in terms of total $\varkappa$ which is the sum of all possible
contributions. 

Within the standard model
the weak nuclear charge $Q_W$ is given by~\cite{SM}
\begin{equation}
Q_W \approx -0.9877N + 0.0716Z.
\end{equation}
Here $N$ is the number of neutrons, $Z$ is the number of protons.

The PNC amplitude of an electric dipole transition between states of
the same parity $|i\rangle$ and $|f \rangle$ is equal to:
\begin{eqnarray}
   E1^{PNC}_{fi}  &=&  \sum_{n} \left[
\frac{\langle f | {\bm d} | n  \rangle
      \langle n | H_{\rm PNC} | i \rangle}{E_i - E_n}\right.
\nonumber \\
      &+&
\left.\frac{\langle f | H_{\rm PNC} | n  \rangle
      \langle n | d_q | i \rangle}{E_f - E_n} \right],
\label{eq:e2}
\end{eqnarray}
where ${\bm d} = -e\sum_i {\bm r_i}$ is the electric dipole operator.
To extract from the measurements the parameter of the nuclear
spin-dependent weak interaction $\varkappa$ one needs to consider PNC
amplitudes between specific hyperfine structure components of the
initial and final states. There amplitudes can be expressed as
\begin{eqnarray}
  E1^{PNC}_{fi} &=&
      (-1)^{F_f-M_f} \left( \begin{array}{ccc}
                           F_f & 1 & F_i  \\
                          -M_f & q & M_i   \\
                           \end{array} \right) \nonumber \\
   &\times& \langle J_f F_f || d_{\rm PNC} || J_i F_i \rangle .
\end{eqnarray}
Here ${\bm F} = {\bm J} + {\bm I}$, $I$ is nuclear spin.
Detailed expressions for the reduced matrix elements of the SI and
SD PNC amplitudes can be found e.g. in Refs.~\cite{Porsev01} and
\cite{JSS03}. For the SI amplitude we have
\begin{eqnarray}
&&\langle J_f,F_f || d_{\rm SI} || J_i,F_i \rangle =
(-1)^{I+F_i+J_f+1}\nonumber \\ 
&& \times \sqrt{(2F_f+1)(2F_i+1)} 
\left\{ \begin{array}{ccc} J_i & J_f & 1 \\
                          F_f & F_i & I \\ 
                    \end{array} \right\}  \label{eq:si0}\\
&&  \times \sum_{n} \left[
\frac{\langle J_f || {\bm d} || n,J_n  \rangle
      \langle n,J_n || H_{\rm SI} || J_i \rangle}{E_i - E_n}\right.  \nonumber \\
&& + \left.\frac{\langle J_f || H_{\rm SI} || n,J_n  \rangle
      \langle n,J_n || {\bm d} || J_i \rangle}{E_f - E_n} \right] \nonumber \\
&& \equiv c(F_f,J_f,F_i,J_i) E^{\prime}_{fi}. \nonumber
\end{eqnarray}
Here $c(F_f,J_f,F_i,J_i)$ is the angular coefficient and the sum over $n$,
$E^{\prime}_{fi}$ does not depend on $F_f$ or $F_i$:
\begin{eqnarray}
  E^{\prime}  &=&  \sum_{n} \left[
\frac{\langle J_f || {\bm d} || n,J_n  \rangle
      \langle n,J_n || H_{\rm SI} || J_i \rangle}{E_i - E_n}\right. \label{eq:si} \\
&+&  \left.\frac{\langle J_f || H_{\rm SI} || n,J_n  \rangle
      \langle n,J_n || {\bm d} || J_i \rangle}{E_f - E_n} \right]. \nonumber 
\end{eqnarray}

For the SD PNC amplitude we have
\begin{eqnarray}
    && \langle J_f,F_f || d_{\rm SD} || J_i,F_i \rangle =
    \frac{G_F}{\sqrt{2}} \varkappa \nonumber \\
     &&\times  \sqrt{(I+1)(2I+1)(2F_i+1)(2F_f+1)/I}  \nonumber \\
    &&\times
     \sum_{n} \left[ (-1)^{J_f - J_i}
     \left\{ \begin{array}{ccc}
     J_n  &  J_i  &   1    \\
      I   &   I   &  F_i   \\                                  
     \end{array} \right\}
     \left\{ \begin{array}{ccc}
      J_n  &  J_f  &  1   \\
      F_f  &  F_i  &  I   \\
     \end{array} \right\} \right. \nonumber \\
  &&\times \frac{ \langle J_f || {\bm d} || n, J_n \rangle
     \langle n, J_n || {\bm \alpha}\rho || J_i \rangle }{E_n -
     E_i} \label{eq:dsd}  \\
  &&+
     (-1)^{F_f - F_i}
     \left\{ \begin{array}{ccc}
     J_n  &  J_f  &   1    \\
      I   &   I   &  F_f   \\
     \end{array} \right\}
     \left\{ \begin{array}{ccc}
     J_n  &  J_i  &  1   \\
     F_i  &  F_f  &  I   \\
     \end{array} \right\} \nonumber \\
 &&\times
     \left. \frac{\langle J_f || {\bm \alpha}\rho ||n,J_n \rangle
            \langle n,J_n || {\bm d} ||J_i \rangle}{E_n - E_f}  \right].
\nonumber
\end{eqnarray}
The PNC amplitude between different hfs components of the
initial and final states can be presented in a form
\begin{equation}
E_{\rm PNC}(F_1,F_2) =c(F_1,F_2)E^{\prime}Q_w\left[1+R(F_1,F_2)\varkappa\right],
\label{eq:EE}
\end{equation}
where $c$  is  an angular coefficient and $R$ is the ratio of
the spin-dependent to the spin-independent PNC amplitudes.

If at least two PNC amplitudes $E_1$ and $E_2$ are measured
then the value of $\varkappa$ can be expressed via the measured ratio of
the  amplitudes $E_1/E_2$ and the calculated ratios $R$ of the SD and SI PNC
amplitudes. 
\begin{eqnarray}
  E_1 = c_1 E^{\prime} Q_W (1 + R_1\varkappa), \nonumber \\
  E_2 = c_2 E^{\prime} Q_W (1 + R_2\varkappa),  \\
  \varkappa = \frac{c_1/c_2 - E_1/E_2}{R_2E_1/E_2 -
      R_1c_1/c_2}.
\label{eq:kappa}
\end{eqnarray}
The ratios $R_1,R_2$ are much less sensitive to numerical
uncertainties than each of the SD and SI PNC amplitudes~\cite{ratio}. 

\section{Calculations}
 
\begin{table}
\caption{Energy levels (in cm$^{-1}$) and $g$-factors of low states of
  mercury. States considered for the PNC transitions are marked as A
  and B.}
\label{t:Hgen}
\begin{ruledtabular}
\begin{tabular}{llcrl rl}
Configu- & & &\multicolumn{2}{c}{Calculations}
&\multicolumn{2}{c}{Experiment} \\  
ration & & State & \multicolumn{1}{c}{Energy} & \multicolumn{1}{c}{$g$} &
\multicolumn{1}{c}{Energy} & \multicolumn{1}{c}{$g$} \\ 
\hline
$6s^2$   &   &  $^1$S$_0$   &      0 &  0.000  &     0 & 0.000 \\ 
$6s6p$ & A$_1$ &  $^3$P$^o_0$ &  38202 &  0.000  & 37645 & 0.000 \\ 
       & A$_2$ &  $^3$P$^o_1$ &  39955 &  1.480  & 39412 & 1.486 \\ 
       & A$_3$ &  $^3$P$^o_2$ &  44812 &  1.500  & 44043 & 1.501 \\ 
         & B  &  $^1$P$^o_1$ &  53584 &  1.020  & 54069 & 1.015  \\ 
$6s7s$   &   &  $^3$S$_1$   &  61879 &  2.000  & 62350 & 2.003 \\ 
         &   &  $^1$S$_0$   &  63399 &  0.000  & 63928 & 0.000 \\ 
\end{tabular}
\end{ruledtabular}
\end{table}

\begin{table}
\caption{Energy levels (in cm$^{-1}$) and $g$-factors of low states of
  xenon. States considered for the PNC transition are marked as A
  and B.}
\label{t:Xeen}
\begin{ruledtabular}
\begin{tabular}{llcrl rl}
Configu- & & &\multicolumn{2}{c}{Calculations}
&\multicolumn{2}{c}{Experiment} \\  
ration & & $J$ & \multicolumn{1}{c}{Energy} & \multicolumn{1}{c}{$g$} &
\multicolumn{1}{c}{Energy} & \multicolumn{1}{c}{$g$} \\ 
\hline
$5p^6$     &  & 0 &      0 &  0.000  &      0  & 0.000 \\ 
$5p^56s$   & A & 2 &  64680 &  1.500  &  67068  & 1.501 \\ 
           &  & 1 &  66089 &  1.214  &  68046  & 1.206 \\ 
           &  & 0 &  74844 &  0.000  &  76197  & 0.000 \\
           & B & 1 &  75999 &  1.314  &  77186  & 1.321 \\ 
$5p^56p$   &  & 1 &  76083 &  1.853  &  77270  & 1.852 \\
           &  & 1 &  77914 &  1.024  &  78957  & 1.022 \\
\end{tabular}
\end{ruledtabular}
\end{table}

Calculations for xenon and mercury were performed with the use of the
configuration interaction (CI) method. We treat Hg as an atom with two
valence electrons and Xe as an atom with six valence
electrons. Calculations for Hg are very similar to what we did before
for Hg~\cite{HgCI} and Yb~\cite{YbPNCt}. We use the $V^{N-2}$
approximation in which initial Hartree-Fock procedure is done for the
Hg$^{2+}$ ion. The complete set of single-electron orbitals is
constructed using the B-spline technique~\cite{JohSap86}. Core-valence
correlations are included by adding the second-order correlation
potential $\hat \Sigma$ to the CI Hamiltonian in the framework of the
CI+MBPT method~\cite{CI+MBPT}. Accuracy for the energies is further
improved by rescaling the core-valence correlation operator $\Sigma$
(see \cite{HgCI,YbPNCt} for details). The rescaling coefficients are
$\lambda_s=0.82$ for $s$-states and $\lambda_p=0.9$ for
$p$-states. The calculated energies and $g$-factors of mercury are
presented in Table~\ref{t:Hgen} together with corresponding
experimental numbers. The $g$-factors are useful for the
identification of states. The accuracy for the calculated energies is
within 1\% for majority  of the  states. It is not perfect in spite of the fitting
because with use only two fitting parameters for all states.

Similar approach for xenon is problematic due to the larger number of
valence electrons. We treat all $5p$ electrons as valence ones, so
that total number of valence electrons is six. Using the same
technique as for Hg would lead to a very large CI matrix. It was
suggested in \cite{XeHg} that the hole-particle formalism can be used for
the calculation of the electron structure of xenon. In this case only
two active particles enter the CI calculations and the calculations
might not be more complicated than for mercury. This assumes that only
single excitations are allowed from the $5p$ subshell. However, in our
experience double excitations are also important. Inclusion of double
excitations within the hole-particle formalism hugely complicates the
problem. Therefore, we use simpler approach. We use standard CI
technique for six valence electrons. Initial Hartree-Fock procedure is
done for neutral xenon (the $V^{N}$ approximation). Single-electron
basis states above the core are calculated as the Hartree-Fock states
in the $V^{N-1}$ potential of the frozen core. Many-electron basis states for
the CI calculations are formed by allowing single and double
excitations from the $5p$ subshell to the states above the core. 

Accurate treatment of the core-valence correlations for xenon within
the CI+MBPT method is problematic due to large contribution of the
subtraction diagrams~\cite{CI+MBPT}. On the other hand the
core-valence correlations are relatively small for xenon compared to
valence-valence correlations due to large number of valence
electrons. Therefore, they can be included approximately.

To simulate the effect of core-valence correlations we include into
the CI Hamiltonian the core polarization potential
\begin{equation}
  \delta V = - \frac{\alpha_c}{a^4+r^4},
\label{eq:pol}
\end{equation}
where $\alpha_c$ is the polarizability of the core and $a$ is a
cut-off parameter. We use $a=a_B$ and treat $\alpha_c$ as a fitting
parameter. This allows us to fit the energy interval $\Delta
E = 84 \ {\rm cm}^{-1}$ between state B and close state of the same
total momentum $J$ and opposite parity belonging to the $5p^56p$
configuration. This is important because PNC amplitude is very
sensitive to this energy interval. We use $\alpha_c=0.554$ a.u. for
$s$ states of valence electrons and $\alpha_c=0$ for other states.

The results for energies and $g$-factors of xenon are presented in
Table~\ref{t:Xeen}. Note that the accuracy is similar to what we have
for mercury. This is because the the core-valence correlations are strong for
mercury, they are significantly stronger than the correlations between
two valence electrons. This is a limitation factor for the accuracy of
the calculation. In contrast, the core-valence correlations are small for
xenon.  

\subsection{Dalgarno-Lewis method for matrix elements}

To calculate PNC amplitude we need to calculate matrix elements
between many-electron states and perform summation over complete set
of many-electron basis states (see, e.g. (\ref{eq:si}) and
(\ref{eq:dsd})). We use random phase approximation (RPA) 
\cite{DzuGin06,DzuFla07} for the matrix elements and Dalgarno-Lewis
method~\cite{DalLew55} for the summation. 

Matrix elements are given by
\begin{equation}
  E1_{vw} = \langle \Psi_v || \hat f + \delta V_f || \Psi_w
  \rangle, \label{E1} 
\end{equation}
where $\delta V$ is the correction to the core potential due to the core
polarization by an external field $\hat f$. In present calculations
$\hat f$ represents either external electric field, SI weak
interaction or SD PNC interaction.

Summation over complete set of many-electron states is reduced to
calculation of the correction $\delta \Psi_v$ to the many-electron
wave function of the state $v$ due to the weak interaction
perturbation $H_{\rm PNC}$. Then, the PNC amplitude is given by
\begin{equation}
  A_{vw} = \langle \delta \Psi_v ||{\bm d}|| \Psi_w \rangle \, .
\label{eq:deltapsi}
\end{equation}
The correction $\delta \Psi_v$ is found by solving the system of
linear inhomogeneous equations
\begin{equation}
  (\hat H^{\rm eff} - E_v )\delta \Psi_v = - (\hat H_{\rm PNC}+\delta
  V_{\rm PNC}) \Psi_v.
\label{eq:DL}
\end{equation}

The proposal of \cite{XeHg} considers the PNC transitions between the 
exited states A and B in Hg and Xe (see Tables \ref{t:Hgen} and
\ref{t:Xeen}). The upper state B in both atoms is very close to
another state of the same total momentum $J$ but opposite parity.  
The interval is 8282 cm$^{-1}$ for Hg and only 84 cm$^{-1}$ in Xe. This
is a strong advantage of using these transitions from the experimental
point of view because the proximity of the sates of the same total angular 
momentum but opposite parity leads to the strong enhancement of the PNC
amplitude. On the other hand, this is a complication from the
theoretical point of view. The PNC amplitudes are sensitive to small
energy intervals where a small error in the calculated energy of the
states can lead to a large error in the value of the PNC amplitude. To
go around this problem we use a stabilizing procedure which consists
of the following steps. First, we use the procedure described above
(see Eq. (\ref{eq:deltapsi}) and (\ref{eq:DL})) without
modifications. Then we repeat the calculations applying the
orthogonality conditions for the $\delta \Psi_{\rm B}$ to the close
state of same $J$ and parity. The contribution of the close state is
found by comparing the two results. Finally, this contribution is
added back to the PNC amplitude with the rescaling parameter $\Delta
E_{\rm theor}/\Delta E_{\rm exp}$. 

The procedure described above corresponds to the exact
fitting of the energy denominators in (\ref{eq:si}) and (\ref{eq:dsd}).
Therefore, the same results should be obtained if the important energy
intervals are fitted exactly by rescaling the correlation potential
$\hat \Sigma$ (for Hg) or polarization potential $\delta V$ (for
Xe). This is another important test which we used in the calculations.

\subsection{Accuracy of the calculations}

\label{s:a}
Accuracy of the similar calculations for two-valence-electron atoms Hg
and Yb were discussed in detail in our earlier
works~\cite{HgCI,YbPNCt,Ybpol}. It was demonstrated that the accuracy
for transition amplitudes and polarizabilities is on the level of
5\%. Note that the PNC amplitude is a second-order effect similar to
polarizability but with one electric dipole operator replaced by weak
interaction. In present work we assume the same 5\% uncertainty for the
PNC effect in mercury. This is supported by the study of the
limitation factors which are discussed below. 

The uncertainty for xenon is higher due to larger number of valence
electrons which make it difficult to saturate the basis. We assume the
10\% uncertainty which come from comparing the results obtained with
two different basis sets, the b-spline basis set and the Hartree-Fock
basis set.

The main factor limiting the accuracy of the calculations of the PNC
amplitudes in xenon and mercury is the proximity of the levels of the
same total momentum $J$ and opposite parity. These states are mixed by
weak interaction and small energy interval between them leads to stong
enhancement of the PNC effect. This is one of the main reason for the
choice of atoms and transitions. However, it represents a challenge for
the calculations. Even small theoretical error in the energies of the
mixed states can lead to large error in the PNC amplitude. There are
two ways around this problem. One is to fit the energy interval
exactly. Another is to isolate the resonant contribution from the rest
of the amplitude and to re-scale it to the correct energy interval. We
did both things for xenon and found that they give the same
results. Therefore, for mercury we just fit the energies. As long as
the problem of small energy intervals is understood and properly dealt
with it didn't contribute much to the uncertainty of the results.

Another important limitation factor for xenon, which is harder to deal
with, is large number of valence electrons. It was suggested in
Ref.~\cite{XeHg} that xenon can be treated as a system with one
electron and one hole in the $5p$ subshell. This would make it a
two-particle system similar to mercury. However, such approach assumes
that only single excitations from the $5p$ subshell are considered. In
our experience full saturation of the basis is not possible without
inclusion of double and may be even triple excitations. This would
lead to the CI matrix of the extremely large size. Probably, this can
be done with the use of supercomputers. Another open question for
xenon is whether and how to include core-valence
correlations. Incomplete saturation of the basis is the main source of
the uncertainty of present calculations for xenon.

Saturation of the basis is not a problem for mercury when it is
considered as a system with two valence electrons. However, here we
have another problem, strong core-valence correlations. The outermost
subshell of the mercury core, $5d$ is strongly mixed with the valence
$6s$ electrons. This is evident from the presence in the discrete
spectrum of mercury the states with excitations for the $5d$
subshell. The core-valence correlations are included in present work
with the use of the second-order correlation operator $\hat
\Sigma$. Most probably the correlations are too strong to be treated
accurately within the second-order approach. The answer may come from
the use of some all-order technique  similar to what was recently
developed in Ref.~\cite{Safronova}.

Even though some improvement of the accuracy of the calculations is
possible, there is little chance that it will ever match the accuracy
for cesium~\cite{CsPNC}. Therefore the main focus of the PNC study in
Xe and Hg should be directed to the measurements of the anapole
moments and to the study of the ratio of PNC effects in different
isotopes. 

\section{Results and Discussion}

\begin{table*}
\caption{PNC amplitudes ($z$-components) for the $|5p^56s^2 \ ^2[3/2]^o_2,F_1
  \rangle \rightarrow  |5p^56s \ ^2[1/2]^o_1,F_2\rangle$ transitions in
  $^{129}$Xe and $^{131}$Xe. $I$ is nuclear spin, ${\bm F} = {\bm J} +
  {\bm I}$. $E^{\prime}$ is given by (\ref{eq:si}).} 
\label{t:Xepnc}
\begin{ruledtabular}
\begin{tabular}{ccccc rr}
$A$ & $I$ & $E^{\prime}$ & $F_1$ & $F_2$ &  \multicolumn{2}{c}{PNC amplitude} \\
    &     & units $10^{-10}iea_0$ &&& \multicolumn{1}{c}{$E^{\prime}Q_W$} & \multicolumn{1}{c}{$10^{-10} iea_0$} \\ 
\hline
129 & 1/2 & 3.16 & 3/2 & 1/2 & $    (1/3)(1+0.0387\varkappa)$ &   $1.05(1+0.0387\varkappa)$ \\
    &     &      & 3/2 & 3/2 & $-1/\sqrt(50)(1+0.010\varkappa)$ & $-0.45 (1+0.010\varkappa)$\\
    &     &      & 5/2 & 3/2 & $\sqrt(2/25)(1-0.0226\varkappa)$ & $0.89 (1-0.0226\varkappa)$\\

131 & 3/2 & 3.25 & 1/2 & 1/2 & $-1/\sqrt{18} (1+0.0345\varkappa)$   & $-0.766 (1+0.0345\varkappa)$  \\ 
    &     &      & 1/2 & 3/2 & $-1/\sqrt{90} (1+0.0252\varkappa)$  & $-0.343(1+0.0252\varkappa)$  \\
    &     &      & 3/2 & 1/2 & $ 1/\sqrt{18} (1+0.0282\varkappa)$ & $ 0.766 (1+0.0282\varkappa)$ \\
    &     &      & 3/2 & 3/2 & $-\sqrt{8/125}(1+0.0189\varkappa)$  & $-0.822 (1+0.0189\varkappa)$ \\ 
    &     &      & 3/2 & 5/2 & $-1/\sqrt{375}(1+0.0034\varkappa)$  & $-0.178(1+0.0034\varkappa)$ \\ 
    &     &      & 5/2 & 3/2 &  $\sqrt{7/125}(1+0.0083\varkappa)$ & $ 0.769 (1+0.0083\varkappa)$ \\
    &     &      & 5/2 & 5/2 & $-\sqrt{3/70} (1-0.0072\varkappa)$  & $-0.673 (1-0.0072\varkappa)$ \\ 
    &     &      & 7/2 & 5/2 &  $\sqrt{2/35} (1-0.0220\varkappa)$  & $ 0.777 (1-0.02204\varkappa)$ \\
\end{tabular}
\end{ruledtabular}
\end{table*}

\begin{table*}
\caption{PNC amplitudes ($z$-components) for the $|6s6p \ ^3$P$^o_J,F_1
  \rangle \rightarrow  |6s6p \ ^1$P$^o_1,F_2\rangle$ transitions in
  $^{199}$Hg and $^{201}$Hg.  $I$ is nuclear spin, ${\bm F} = {\bm J} +
  {\bm I}$. $E^{\prime}$ is given by (\ref{eq:si}).} 
\label{t:Hgpnc}
\begin{ruledtabular}
\begin{tabular}{cccccc rr}
$A$ & $I$ & $J$ & $E^{\prime}$ & $F_1$ & $F_2$ &  \multicolumn{2}{c}{PNC amplitude} \\
    &    &&units $10^{-10}iea_0$ &&& \multicolumn{1}{c}{$E^{\prime}Q_W$} & \multicolumn{1}{c}{$10^{-10} iea_0$} \\ 
\hline
199 & 1/2 &  0 &  3.41 & 1/2 &  1/2 &       $(1/3) (1+0.0084\varkappa)$ &   $1.14 (1+0.0084\varkappa)$ \\
    &     &    &       & 1/2 &  3/2 &  $-\sqrt{2/9} (1-0.0042\varkappa)$ &  $-1.61 (1-0.0042\varkappa)$ \\
    &     &  1 &  5.31 & 1/2 &  1/2 &  $ \sqrt{2/27}(1+0.0538\varkappa)$ &  $ 1.44 (1+0.0538\varkappa)$ \\
    &     &    &       & 1/2 &  3/2 &   $\sqrt{1/27}(1-0.0860\varkappa)$ &   $1.02 (1-0.0860\varkappa)$ \\
    &     &    &       & 3/2 &  1/2 &   $\sqrt{1/27}(1+0.0308\varkappa)$ &   $1.02 (1+0.0308\varkappa)$ \\
    &     &    &       & 3/2 &  3/2 &    $1/\sqrt{6}(1-0.0105\varkappa)$ &   $2.17 (1-0.0105\varkappa)$ \\
    &     &  2 &  3.71 & 3/2 &  1/2 &    $   (1/3) (1+0.0381\varkappa)$ &   $1.24 (1+0.0381\varkappa)$ \\
    &     &    &       & 3/2 &  3/2 &  $-1/\sqrt{50}(1+0.0471\varkappa)$ &  $-0.525 (1+0.0471\varkappa)$ \\
    &     &    &       & 5/2 &  3/2 &  $ \sqrt{2/25}(1-0.0264\varkappa)$ &  $ 1.05  (1-0.0264\varkappa)$ \\
201 & 3/2 &  0 &  3.47 & 3/2 &  1/2 &    $   (1/3) (1+0.0069\varkappa)$ &   $1.16 (1+0.0069\varkappa)$ \\
    &     &    &       & 3/2 &  3/2 &    $1/\sqrt{5}(1+0.0028\varkappa)$ &   $1.55 (1+0.0028\varkappa)$ \\
    &     &    &       & 3/2 &  5/2 &  $-\sqrt{2/15}(1-0.0041\varkappa)$ &  $-1.27 (1-0.0041\varkappa)$ \\
    &     &  1 &  5.40 & 1/2 &  1/2 &  $-1/\sqrt{54}(1+0.0171\varkappa)$ &  $-0.735 (1+0.0171\varkappa)$ \\
    &     &    &       & 1/2 &  3/2 &   $\sqrt{5/54}(1+0.00362\varkappa)$& $1.64 (1+0.00362\varkappa)$ \\
    &     &    &       & 3/2 &  1/2 &   $\sqrt{5/54}(1+0.0419\varkappa)$ &   $1.64 (1+0.0419\varkappa)$ \\
    &     &    &       & 3/2 &  3/2 &   $\sqrt{2/75}(1+0.0607\varkappa)$ &   $0.882 (1+0.0607\varkappa)$ \\
    &     &    &       & 3/2 &  5/2 &   $     (1/5)(1-0.0515\varkappa)$ &   $1.08 (1-0.0515\varkappa)$ \\
    &     &    &       & 5/2 &  3/2 &    $    (1/5)(1+0.0122\varkappa)$ &   $1.08 (1+0.0122\varkappa)$ \\
    &     &    &       & 5/2 &  5/2 &    $1/\sqrt{6}(1-0.0103\varkappa)$ &   $2.20 (1-0.0103\varkappa)$ \\
    &     &  2 &  3.78 & 1/2 &  1/2 &  $-1/\sqrt{18}(1+0.0385\varkappa)$ &  $-0.890 (1+0.0385\varkappa)$ \\
    &     &    &       & 1/2 &  3/2 &  $-1/\sqrt{90}(1+0.0414\varkappa)$ &  $-0.398 (1+0.0414\varkappa)$ \\
    &     &    &       & 3/2 &  1/2 &   $1/\sqrt{18}(1+0.0240\varkappa)$ &   $0.890 (1+0.0240\varkappa)$ \\
    &     &    &       & 3/2 &  3/2 & $-\sqrt{8/125}(1+0.0269\varkappa)$ &  $-0.956 (1+0.0269\varkappa)$ \\
    &     &    &       & 3/2 &  5/2 & $-1/\sqrt{375}(1+0.0318\varkappa)$ &  $-0.195 (1+0.0318\varkappa)$ \\
    &     &    &       & 5/2 &  3/2 &  $\sqrt{7/125}(1+0.00286\varkappa)$&  $ 0.894 (1+0.00286\varkappa)$\\
    &     &    &       & 5/2 &  5/2 &  $-\sqrt{3/70}(1+0.00774\varkappa)$&  $-0.782 (1+0.00774\varkappa)$\\
    &     &    &       & 7/2 &  5/2 &  $ \sqrt{2/35}(1-0.0260\varkappa)$ &  $ 0.903 (1-0.0260\varkappa)$ \\
\end{tabular}
\end{ruledtabular}
\end{table*}

The calculated nuclear spin-independent PNC amplitude for Xe
($z$-component) is 
\begin{equation} 
  E_{\rm PNC}({\rm A} \rightarrow {\rm B}) = 1.76 \times 10^{-10}
  (-Q_W/N) iea_B.
\label{eq:XePNC}
\end{equation}
The spin-independent PNC amplitudes for Hg are
\begin{eqnarray} 
  E_{\rm PNC}({\rm A_1} \rightarrow {\rm B}) &=& 2.09 \times 10^{-10}
  (-Q_W/N) iea_B, \nonumber \\
  E_{\rm PNC}({\rm A_2} \rightarrow {\rm B}) &=& 1.77 \times 10^{-10}
  (-Q_W/N) iea_B, \label{eq:HgPNC} \\
  E_{\rm PNC}({\rm A_3} \rightarrow {\rm B}) &=& 1.25 \times 10^{-10}
  (-Q_W/N) iea_B. \nonumber
\end{eqnarray}
The difference in the value of the PNC amplitude for different
isotopes is mostly due to different value of the weak nuclear charge
$Q_W$. Therefore, the amplitudes (\ref{eq:XePNC}) and (\ref{eq:HgPNC})
may be used for any isotope.

Detailed data for both SD and SI PNC amplitudes for isotopes with
non-zero nuclear spin are presented in Table \ref{t:Xepnc} for Xe and Table
\ref{t:Hgpnc} for Hg.

\subsection{M1 amplitudes}

The experimental proposal \cite{XeHg} is aimed to measure the PNC
optical rotation. The angle of rotation is proportional to the ratio
$R = Im(E_{\rm PNC}/M1)$. Therefore, we need to know the values of the
magnetic dipole amplitudes for the transitions proposed for PNC
measurements. The most accurate values of the M1 amplitudes can be
found analytically using experimental values of the magnetic
$g$-factors to find the coefficients for configuration mixing. 

This is especially important for the case of mercury where numerical
calculations of the M1 amplitudes give unstable results. The reason
for this instability is easy to understand. The transitions considered
for the PNC measurements in mercury are between states of different
spin ($S=1$ for states A$_{1,2,3}$ and $S=0$ for state B). This means
that the M1 amplitudes between these states vanish in the non-relativistic
limit. In relativistic calculations the amplitudes are not zero but
small. These small values are obtained as a result of strong
cancelation between different contributions. This stong cancelation
leads to unstable results.

On the other hand analytical evaluation of the M1 amplitudes is simple
and produce very accurate results. The operator of the magnetic dipole
transition ($M_z = (L_z +2S_z)\mu_B$) has no radial part and cannot
change a principal quantum number in the non-relativistic limit. Therefore, the magnetic $g$-factors
and M1 amplitudes  are mainly  sensitive  
to the mixing of the states belonging to the same configuration. Mixing with other configurations normally produces corrections at the $10^{-3}$ level \cite{g}. We may see this  in Table
\ref{t:Hgen} and \ref{t:Xeen} where $g$-factors of "pure" $sp$ states with $J=2$ and sum of the $g$-factors for mixed states $J=1$ differ from the experimental values by less than 0.1 \%.
Therefore, the mixing coefficients
for  the  states belonging to the same configuration and M1 amplitudes can be
found practically exactly from the known values of the $g$-factors.  Note that we use the calculated value of the overlap between the radial wave functions $p_{1/2}$ and $p_{3/2}$ which is close but not equal to 1. For $Hg$ it is 0.988.

We get for the states A and B of xenon
\begin{eqnarray}
&&  \Psi_A = |5p_{3/2}6s\rangle \nonumber \\
&&  \Psi_B = 0.05|5p_{3/2}6s\rangle + 0.999|5p_{1/2}6s\rangle.
\label{eq:Bxe}
\end{eqnarray}
This leads to the M1 amplitude
\begin{equation}
M1_{AB} = 1.22\mu_B = 0.00446 ea_B.
\label{eq:M1Xe}
\end{equation}
Using $E^{\prime}$ from table \ref{t:Xepnc} we get $R=7.1(7)(35) \times
10^{-8}$ for $^{129}$Xe and $R=7.3(7) \times
10^{-8}$ for $^{131}$Xe. Here we assume the 10\% uncertainty as it has 
been discussed above. These values of $M1$ and $R$ are close but not in
perfect agreement to what was found in Ref.~\cite{XeHg}: $M1=0.0042
ea_B$, $R=11(3) \times 10^{-8}$. The reason for difference in M1 is not
clear. The authors of Ref.~\cite{XeHg} use slightly different
coefficients of configuration mixing in (\ref{eq:Bxe}). Their values
are 0.062 and 0.998. However, if we use these coefficients we get
$M1=0.00444 ea_B$ which is  different from the value $M1=0.0042
ea_B$ presented in Ref.~\cite{XeHg}.

The situation is more complicated for mercury. All the M1 transitions
of interest happen between the states with different total spin and
vanish in the non-relativistic limit since the operator   $L_z +2S_z$
conserves the total spin.  This leads to the strong suppression. The
values of the M1 
amplitudes presented in Ref.~\cite{XeHg} are too large for the
spin-forbidden transitions. 
No choice of the configuration mixing coefficients can reproduce them. 

The wave functions for states A$_{1,2,3}$ and B for Hg have the form
\begin{eqnarray}
&&  \Psi_{\rm A_1} = |6p_{1/2}6s\rangle \nonumber \\
&&  \Psi_{\rm A_2} = 0.432|6p_{3/2}6s\rangle - 0.902|6p_{1/2}6s\rangle,
\nonumber \\
&&  \Psi_{\rm A_3} = |6p_{3/2}6s\rangle. \label{eq:Bhg}  \\ 
&&  \Psi_{\rm B} = 0.902|6p_{3/2}6s\rangle +
0.432|6p_{1/2}6s\rangle, \nonumber
\end{eqnarray}
The coefficients 0.902 and 0.432 are chosen to fit the experimental
$g$-factors of the states A$_2$ and B.
When projections are included all $|^{2S+1}P_{JJ_z}\rangle$ states of the
$6s6p$ configuration can be written as
\begin{eqnarray}
 |{\rm A_1}: \ ^3P^o_{00}\rangle &=& -\frac{1}{\sqrt{2}}\left[
  |6s_{\frac{1}{2}\frac{1}{2}}6p_{\frac{1}{2}-\frac{1}{2}}\rangle
  -|6s_{\frac{1}{2}-\frac{1}{2}}6p_{\frac{1}{2}\frac{1}{2}}\rangle
\right], \nonumber \\
|{\rm A_2}: \ ^3P^o_{11}\rangle &=& 
  -0.216|6s_{\frac{1}{2}\frac{1}{2}}6p_{\frac{3}{2}\frac{1}{2}}\rangle+ \nonumber \\
&&  0.374|6s_{\frac{1}{2}-\frac{1}{2}}6p_{\frac{3}{2}\frac{3}{2}}\rangle
  +0.902|6s_{\frac{1}{2}\frac{1}{2}}6p_{\frac{1}{2}\frac{1}{2}}\rangle,
  \nonumber \\
 |{\rm A_3}: \ ^3P^o_{22}\rangle &=&
|6s_{\frac{1}{2}\frac{1}{2}}6p_{\frac{3}{2}\frac{3}{2}}\rangle, 
\label{eq:6s6pm} \\
|{\rm B \ }: \ ^1P^o_{11}\rangle &=& 
  -0.451|6s_{\frac{1}{2}\frac{1}{2}}6p_{\frac{3}{2}\frac{1}{2}}\rangle+ 
  \nonumber \\
&&  0.781|6s_{\frac{1}{2}-\frac{1}{2}}6p_{\frac{3}{2}\frac{3}{2}}\rangle
  -0.432|6s_{\frac{1}{2}\frac{1}{2}}6p_{\frac{1}{2}\frac{1}{2}}\rangle.
   \nonumber 
\end{eqnarray}
Here we use (\ref{eq:Bhg}) for the expansion. The results for M1
amplitudes obtained with the use of these formulas are
presented in Table~\ref{t:M1} together with 
the values from Ref.~\cite{XeHg}. Note that there must be $M1 \ll
\mu_B$ for the spin-forbidden transitions. This holds for the results
of present work but not for the results of Ref.~\cite{XeHg}.
Table~\ref{t:M1} also presents the PNC amplitudes and the ratios
$R=Im(E_{\rm PNC}/M1)$. The numbers include 5\% error bars according
to the estimated uncertainty of the calculations which was discussed
in section \ref{s:a}. Note that the values of the M1 amplitudes
obtained in the present work are practically exact due to the fitting of
the experimental $g$-factors.

The values of the ratios $R$ are larger than in
Ref.~\cite{XeHg} due to smaller M1 amplitudes. The values of
$R$ for mercury are about an order on magnitude larger than for xenon (see
above) and about an order of magnitude larger than for Tl, Pb and
Bi~\cite{TlPNC,PbPNC,BiPNC}.  

\begin{table*}
\caption{Magnitudes of the M1 and PNC amplitudes (reduced matrix elements) and ratios $R
  \equiv Im(E_{\rm PNC}/M1)$ for the $^3P_J - 
  ^1P_1$ transitions in mercury. Values for M1 from Ref.~\cite{XeHg}
  are also presented for comparison.} 
\label{t:M1}
\begin{ruledtabular}
\begin{tabular}{l cccc cccc}
Transition & \multicolumn{4}{c}{M1 amplitudes} &
\multicolumn{2}{c}{$Im(E_{\rm PNC})$}  & \multicolumn{2}{c}{$R$} \\ 
           & \multicolumn{2}{c}{This work} &
           \multicolumn{2}{c}{Ref.~\cite{XeHg}} & $^{199}$Hg & $^{201}$Hg & $^{199}$Hg & $^{201}$Hg \\
           &units  $10^{-4}ea_B$ & $\mu_B$ & $10^{-4}ea_B$ & $\mu_B$ &
           \multicolumn{2}{c}{$10^{-10}ea_B$} & \multicolumn{2}{c}{$10^{-7}$} \\
\hline
$^3P^o_0 - ^1P^o_1$ & 8.37 & 0.229 & 14 & 0.384 & 3.4(2) & 3.5(2) & 4.1(2)& 4.2(2) \\
$^3P^o_1 - ^1P^o_1$ & 7.26 & 0.199 & 42 & 1.15  & 5.3(3) & 5.4(3) & 7.3(4)& 7.4(4) \\
$^3P^o_2 - ^1P^o_1$ & 9.94 & 0.272 & 57 & 1.56  & 3.7(2) & 3.8(2) & 3.7(2)& 3.8(2) \\
\end{tabular}
\end{ruledtabular}
\end{table*}

\subsection{Optical rotation}

The angle of optical rotation is given by~\cite{Khrip}
\begin{equation}
  \phi_{\rm PNC} =-\frac{4\pi l}{\lambda}\left( n(\omega)-1\right)R,
\label{eq:phi}
\end{equation}
where $l$ is path length in vapor, $\lambda$ is wavelength of laser
light, $\omega$ is its frequency, and $n(\omega)$ is refractive index
due to the absorption. Although the angle is proportional to the ratio
$R=Im(E_{\rm PNC}/M1)$, the small values for the M1 amplitudes do not
necessarily translate into a large angle of rotation. This is because the
refractive index also depends on M1 amplitude, $n(\omega)-1 \sim
|M1|^2$, which leads to $\phi \sim M1 \cdot E_{\rm PNC}$. However, the
suppression due to the small M1 amplitude can be compensated at
sufficiently high vapor pressure by the appropriate choice of the path
length $l$. These questions are discussed in detail in
Ref.~\cite{Khrip}. Here we just note that the angle of rotation per
unit length is  $\phi \sim M1 \cdot E_{\rm PNC}$. However, the angle of
rotation per absorption length $\phi \sim R$ and it is large for small M1. 

\acknowledgements

The work was supported by the Australian Research Council.
The authors are grateful to D. Budker and T. P. Rakitzis for stimulating 
discussions.

\end{document}